\documentclass[preprint,12pt]{elsarticle}

\usepackage{amsmath,amssymb,amsfonts,amsthm}
\usepackage{comment}
\theoremstyle{plain}
\newtheorem{theorem}{Theorem}

\newtheorem{lemma}[theorem]{Lemma}

\theoremstyle{definition}
\newtheorem{definition}[theorem]{Definition}
\newtheorem{example}[theorem]{Example}

\theoremstyle{remark}

\usepackage{xspace}

\newcommand{\ecalculus}{$\forall$\textsf{Exp+Res}\xspace}
\newcommand{\irc}{\textsf{IR-calc}\xspace}

\newcommand{\qrc}{\textsf{Q-Res}\xspace}

\newcommand{\qrat}{\textsf{QRAT}\xspace}

\newcommand{\lqrc}{\textsf{LD-Q-Res}\xspace}

\usepackage{tikz}
\usetikzlibrary{shapes,snakes}

\usetikzlibrary{fadings}
\tikzstyle{uedge}=[draw=blue!50!red]
\tikzstyle{fedge}=[draw=blue]
\tikzstyle{iedge}=[draw=red]
\tikzstyle{redge}=[draw=green!50!black]
\tikzstyle{rnode}=[draw,inner sep=2pt,color=black]
\tikzstyle{tnode}=[circle,minimum width=3pt,fill,inner sep=0pt]
\tikzstyle{dotnode}=[circle,minimum width=2pt,fill,inner sep=0pt]
\tikzstyle{labn}=[font=\sffamily,circle,fill=white,inner sep=1pt,draw=black]
\tikzstyle{legn}=[font=\scriptsize]
\tikzstyle{reln}=[circle,fill=white,inner sep=.4pt,draw=black]
\tikzstyle{oreln}=[circle,fill=white,inner sep=.4pt,draw=black!50,solid]
\tikzstyle{oree}=[thick,draw=black!50,densely dashed]
\tikzstyle{ree}=[thick,draw=black]
\tikzstyle{calcn}=[rectangle%
                   ,rounded corners=2mm%
                   ,thick%
                   ,draw=black%
                   ,top color=white,bottom color=black!20,%
                   minimum height=.5cm,minimum width=.5cm,inner sep=2pt%
                 ]
\tikzstyle{expcalcn}=[rectangle%
                   ,thick%
                   ,draw=green!40!black
                   ,top color=white,bottom color=green!20,%
                   minimum height=.5cm,minimum width=.5cm,inner sep=2pt%
                 ]
\usepackage{caption}
\usepackage{subcaption}
\begin{document}

\begin{frontmatter}
\title{Does \qrat simulate \irc? \qrat simulation algorithm for \ecalculus cannot be lifted to \irc }
\author[iitropar]{Sravanthi Chede}
\author[iitropar]{Anil Shukla}
\address[iitropar]{Department of Computer Science and Engineering, IIT Ropar, India. {\sf \{sravanthi.20csz0001,anilshukla\}@iitrpr.ac.in} }
\begin{abstract}
We show that the \qrat simulation algorithm of \ecalculus from [B.~Kiesl and M.~Seidl, 2019] cannot be lifted to \irc.
\end{abstract}


\end{frontmatter}

\section{Introduction}
Quantified Boolean formulas (QBFs) extend propositional formulas by adding quantification $\exists$ (there exists) and $\forall$ (for all) to the variables. Several QBF proof systems like \qrc~\cite{qres_paper}, \lqrc~\cite{Balabanov12}, \ecalculus~\cite{exp_paper}, \irc~\cite{irc_paper} have been developed. However, these proof systems are unable to simulate the preprocessing steps used by several QBF-solvers. To overcome this a new proof system Quantified Resolution Asymmetric Tautologies (\qrat)~\cite{qrat_paper} has been developed and also shown that it is capable of simulating all the existing preprocessing steps used by the current QBF-solvers~\cite{qrat_paper}.

Recently it has been shown that \qrat can even simulate \ecalculus~\cite{KieslS19} and \lqrc~\cite{ldqres_paper}. We know that \irc and \lqrc are incomparable~\cite{ld_simulation} and since \qrat can simulates \lqrc, it implies that \irc cannot simulate \qrat. But, it is still open whether \qrat can simulate \irc?

Since \irc is an extension of \ecalculus, it is very natural to use the \qrat simulation algorithm of \ecalculus for the \irc. In this note we show that this is not possible. That is, we cannot lift the \qrat simulation algorithm of \ecalculus for \irc in general. For proving the same we consider an important family of false QBFs $\phi_n$ from~\cite{exp_paper}, which is known to be easy for \irc and hard for \ecalculus and we show that \irc proof of $\phi_n$ cannot be simulated by the proposed modified algorithm (Section~\ref{sec:modified-algo}) which is the only approach to lift the existing simulation algorithm. 

\section{Definitions}
In this note we assume that QBFs are in closed prenex form i.e., we consider the form $Q_1 X_1 . . .Q_k X_k$.$\psi$, where $X_i$ are pairwise disjoint sets of variables; $Q_i$ $\in$ \{$ \exists$, $ \forall$\} and $Q_i \neq Q_{i+1}$. The propositional part $\psi$ of a QBF is called the matrix which should be in CNF (Conjunctive Normal Form) and the rest is the prefix $Q$. A clause is a disjunction of literals and a CNF is a conjunction of clauses. We denote the empty clause by $\bot$.
If a variable $x$ is in the set $X_i$, we say that $x$ is at level $i$ and write $lv(x)=i$. Note that only variables are in prefix, so given a literal $l$ if $x$=var($l$),then quantifier of $l$ is same as that of $x$.
Given a literal $\ell$ with quantifier $Q_i$ and a literal $k$ with quantifier $Q_j$, we write $\ell  {\leq}_Q k$ if $i \leq j$ (we say that $\ell$ occurs left of $k$).

Informally, a proof system is a function $f$ which maps proofs to theorems (or contradictions). A proof system $f$ simulates another proof system $g$ (i.e., $f \leq_p g$) if every $g$-proof of a theorem (or contradiction) can be efficiently translated into an $f$-proof of the same theorem. Proof systems $f$ and $g$ are said to be incomparable, if none of them can simulate the other.

One of the main approach to QBF-solving is through expansion of quantifiers. 
Several expansion-based QBF proof systems have been developed, for example, \ecalculus~\cite{exp_paper}.
This calculus downloads the axioms by dropping all universal literals in a clause and annotating the existential literals by an assignment to all universal variables which occur to left of that variable.
It also allows the following resolution step: $\frac{(C_1 \lor x^{\tau})~~~~~ (C_2 \lor \overline{x}^{\tau}))}{(C_1 \lor C_2)}$, where $C_1$ and $C_2$ are clauses, $x^\tau$ is a literal, and ($C_1 \vee C_2$) is the resolvent.

The \irc proof system~\cite{irc_paper} has been developed as an extension of the \ecalculus.
Here, in the axiom steps, the existential variables are only annotated with $\forall$ variables which are on its left and belong to the same clause. The following instantiation step is also introduced: 
$inst(\sigma,C)= \{ x^{\tau[\sigma]}~|~x^{\tau} \in C \}$
where, $\sigma$ is a partial assignment to the universal variables and for every $\forall$ variable $\ell$ to the left of x, 
$\tau[\sigma]$ returns $\tau (\ell)$ if $\ell \in dom(\tau)$ else $\sigma (\ell)$ if $\ell \in dom(\sigma)$.
The resolution step remains the same.\\ 

\textbf{\qrat Proof System~\cite{qrat_paper}}: 
We need the following definitions:   
\vspace{-0.2cm}
\begin{definition}[\cite{qrat_paper}]\label{asymmetric-tautology}
Clause $C$ is an \textbf{Asymmetric Tautology} (AT) w.r.t. to CNF $\psi$ iff $\psi$ $\vdash_1  C$. (Alternatively can be checked if $\bot \in$ unit-propagation$(\psi \cup \overline{C}))$. 
Unit propagation($\vdash_1$) simplies a CNF by repeating the following: If there is a unit clause ($\ell$) then remove all clauses that contain the literal $\ell$ and remove the literal $\overline{\ell}$ from all clauses.
\end{definition}
\vspace{-0.2cm}
Given two clauses $(C \lor \ell), (D \lor \overline{\ell})$ of a QBF $Q.\psi$, the outer resolvent OR($Q$,$C$,$D$,$\ell$) is the clause consisting of all literals in $C$ together with those literals of $D$ that occur to the left of $\ell$, i.e $C \cup \{ k~|~k \in D , k \leq_Q \ell \}$. A clause $C\vee \ell$ is \qrat-clause w.r.t a QBF $Q.\psi$ if for every $D\vee \overline{\ell} \in \psi$ the OR($Q$,$C$,$D$,$\ell$) is implied by unit propagation. We say that $\ell$ is the \qrat-literal.

If a clause $C$ contains an existential \qrat literal, it has been shown in \cite{qrat_paper} that 
$C$ can be removed or added without effecting the satisfiability. Also, if a clause $C$ contains a universal \qrat literal $\ell$ then dropping $\ell$ from $C$ is also a satisfiability preserving step. Note that a clause which is AT is also a \qrat-clause on any literal. Additionally, \qrat allows elimination of any clause at any point in the proof. The only remaining rule of \qrat system is the Extended Universal Reduction (EUR) rule. We need the following:
\vspace{-0.2cm}
\begin{definition}[\cite{KieslS19}]\label{resolution-path} 
Given a QBF $\phi$=$Q.\psi$, a universal literal $u$ and an existential literal $e_n$ to the right of $u$, we say that $\phi$ contains a {\bf resolution path} from $u$ to $e_n$ if there exists a sequence $C_1,...,C_n$ of clauses in $\psi$ such that $u \in C_1$, $e_n \in C_n$ along with a sequence $e_1,...,e_{n-1}$ of existential literals which occur to the right of $u$, where $e_i \in C_i$, $\overline{e_i} \in C_{i+1}$ and $var(e_i) \neq var(e_{i+1})$.

The \textbf{reflexive-resolution-path dependency} scheme ($D^{rrs}$) defines that an existential literal $e$ depends on a universal literal $u$ iff one of the following conditions holds: (1) There exist resolution paths from $u$ to $e$ and from $\overline{u}$ to $\overline{e}$. (2) There exist resolution paths from $u$ to $\overline{e}$ and from $\overline{u}$ to $e$.
\end{definition}
\vspace{-0.2cm}
The EUR rule of \qrat allows to remove a universal literal $u$ from a clause $(C \lor u) \in \psi$ where all $\ell \in C$ are independent of $u$ according to $D^{rrs}$.
\section{A brief recap of the \qrat simulation of \ecalculus from~\cite{KieslS19}}
The algorithm from~\cite{KieslS19} starts with a QBF $Q.\psi$ and a \ecalculus proof $\pi$ of $Q.\psi$. Then constructs a \qrat proof $\Pi$ of $Q.\psi$  as follows:\\
\textbf{Step 1 (Introduction of definitions):} For each annotated variable $x^\tau$ in $\pi$, introduce the definition clauses $(\overline{x}^\tau \lor  x)$ and $(x^\tau \lor \overline{x})$ in this order and place $x^\tau$ in the same quantifier block as $x$. Denote the resulting accumulated formula by $Q^{'}.\psi_1 $.
Observe that this step is valid: the definition clause $(\overline{x}^\tau \lor  x)$ is a \qrat since $x^\tau$ is new. Then, the definition $(x^\tau \lor \overline{x})$ is a \qrat since the only outer-resolvent upon $x^\tau$ is the tautology $(\overline{x} \lor  x)$.\\
\textbf{Step 2 (Introduction of annotated clauses):} For each clause $C^\tau \in \pi$ that was obtained from clause $C \in \psi$ by axiom rule, add a clause $C^\tau \lor u_1 \lor ... \lor u_k$ (where $u_1,...,u_k$ are universal literals in $C$). Denote resulting accumulated formula by $Q^{'}.\psi_2 $.
Observe that this new clause is AT w.r.t $\psi_1$.\\
\textbf{Step 3 (Elimination of Input Clauses and Definitions):} $Q^{'}.\psi_3$ =$\psi_2$-$\psi_1$.\\
\textbf{Step 4 (Removal of all universal literals):} Apply EUR rule in annotated clauses for dropping all $\forall$ variables from right to left in the prefix $Q^{'}$.
This completes the axiom steps. This step is valid because of the following Lemma.
\vspace{-0.6cm}
\begin{lemma}[\cite{KieslS19}]\label{lemma1}
If $Q^{'}.\psi_3 $ contains a resolution path from $u$ to $e$, then $e$ must be an annotated literal of the form $l^\tau$ where the assignment $\tau$ falsifies $u$.
\end{lemma}
\vspace{-0.2cm}
This lemma being true implies that $D^{rrs}$ can never be found between $u$ and $e$ as the path sees the literals $u$,$e$ and $\overline{e}$ but never includes $\overline{u}$.This being true for any such e implies $u$ can be safely dropped by EUR rule.\\
\textbf{Step 5 (Resolution proof):} Simulate the remaining resolution steps.
\vspace{-0.2cm}
\subsection{Problems with direct usage of above Algorithm for \irc proofs}
In order to simulate an \irc proof, one also need to simulate the instantiation steps. We observe that if we do not delete all the definition clauses in Step $3$, we can simulate the instantiation steps as well. Note that we may need to introduce more definition clauses for fresh variables when introduced. 
\vspace{-0.7cm}
\begin{lemma}\label{lemma2}
Instantiation step can always be simulated when retaining all the definition clauses in the QBF.
\end{lemma}
\vspace{-0.5cm}
\begin{proof}
Suppose we have a clause $C_i =({x_1}^{\tau_1} \lor {x_2}^{\tau_2} \lor {e_i}^{\tau})$ and the \irc proof applies an instantiation step on $C_i$, i.e, $inst(\sigma,C_i)$. Let the step return ${C_i}^{'} =({x_1}^{\tau_1} \lor {x_2}^{\tau_2 [\sigma]} \lor {e_i}^{\tau [\sigma]})$ and say annotations of $x_1$ have not changed but those of $x_2$ have changed and resulted in a new variable (${x_2}^{\tau_2 [\sigma]}$) not present in the QBF currently and that we wanted to resolve on $e_i$ so it's annotations have changed but the new variable is already existing in the QBF.

${x_2}^{\tau_2 [\sigma]}$ being a new variable, we add its definitions $(\overline{x_2} \lor {x_2}^{\tau_2 [\sigma]}) \land (x_2 \lor \overline{{x_2}}^{\tau_2 [\sigma]})$ and because ${x_2}^{\tau_2}$,${e_i}^{\tau}$,${e_i}^{\tau [\sigma]}$ are already existing we would have $(x_2 \lor \overline{{x_2}}^{\tau_2})$, $(e_i \lor \overline{{e_i}}^{\tau})$, $(\overline{e_i} \lor {e_i}^{\tau [\sigma]})$ clauses already present in the QBF. Observe that a series of resolving steps on these clauses derives the clause ${C_i}^{'}$. Therefore, it is an AT and hence a \qrat-clause and can be added.
\end{proof}
\vspace{-0.2cm}
However, if we retain all the definition clauses, then Lemma~\ref{lemma1} may not always hold, implying that we cannot simulate the axiom download steps. Also we cannot add definition clauses of existing variables after axiom step as they are no longer \qrat clauses.
The only way to lift the algorithm for \irc is to retain all the {\bf important} definition clauses and delete the unimportant ones. We say that a definition clause is important if we need the same for the simulation of an instantiation step later in the \irc proof. 

This motivates us to design a two pass algorithm. In the first pass, the algorithm marks all the important definition clauses and deletes the unimportant ones. The algorithm then checks whether Lemma~\ref{lemma1} holds with all the important clauses present. If no, the algorithm stops. Otherwise, in the second pass the algorithm continues with the successful simulation of the \irc proof. Next, we present the algorithm in detail.
\section{Modified \qrat simulation algorithm for \irc}\label{sec:modified-algo}
The Modified algorithm starts with a QBF $Q.\psi$ and an \irc proof $\pi$ of $Q.\psi$, an constructs a \qrat proof $\Pi$ of $Q.\psi$  as follows: \\
\textbf{Step 1 (Introduction of definition clauses):} In the first pass, we add definition clauses of all annotated variables in the Axiom clauses as defined in the above algorithm. Denote the resulting accumulated formula by $Q^{'}.\psi_1 $. Also give labels to all definition clauses, say $D_1,...,D_{2k}$.\\
\textbf{Step 2 (Introduction of annotated clauses):} Exactly as defined in the Step 2 of above algorithm. Denote the resulting formula by $Q^{'}.\psi_2 $.\\
\textbf{Step 3 (Elimination of input clauses):} We only drop the input clauses from $\psi$ in this step. Denote the resulting accumulated formula by $Q^{'}.\psi_3$.\\
\textbf{Step 4 (Find all important definition clauses):} Assume that the axiom downloads have been performed and go ahead in the \irc proof $\pi$ scanning for instantiation steps. Let $C_i = ({x_1}^{\tau_1} \lor {x_2}^{\tau_2} \lor {x_3}^{\tau_3})$ be any derived clause in $\pi$ and $C_{i+1}$ = inst($\sigma$,$C_i$) be an instantiation step. Say $C_{i+1} = ({x_1}^{\tau_1 [\sigma]} \lor {x_2}^{\tau_2 [\sigma]} \lor {x_3}^{\tau_3})$, where $x_1^{\tau_1 [\sigma]}$ is the literal with modified annotations which already exists in the QBF, $x_2^{\tau_2 [\sigma]}$ is a new variable not yet present in the QBF and $x_3^{\tau_3}$ is the literal whose annotations did not change after the instantiation step (zero or more of each type of variables are allowed in the clause $C_i$).

For each existing changed literal (i.e $x_1$) we note the clause needed for this change in the annotations of $x_1$ i.e $(\overline{x_1}^{\tau_1} \lor {x_1}^{\tau_1 [\sigma]})$. Re-write these clauses in terms of definition clauses i.e $(x_1 \lor \overline{x_1}^{\tau_1}) \land (\overline{x_1} \lor {x_1}^{\tau_1 [\sigma]})$. We mark these definition clauses as important. 
Then, for each new changed literal (i.e $x_2$), we mark the definition clause ($x_2 \lor \overline{x_2}^{\tau_2} $) as important. Note that the other clause needed in this case will be added at the time of actual simulation.\\
\textbf{Step 5 (Drop all unimportant definition clauses added in Step 1).}\\
\textbf{Step 6 (Find resolution paths):} At this point we know that all instantiation steps of $\pi$ can be simulated. The algorithm now checks whether the EUR steps are still applicable to complete the simulation of the axiom steps. We check the same as follows: for every universal variable ($u$) going from right to left order in the prefix $Q^{'}$, check if there exists a resolution path that starts from a clause containing $u$ to one containing $\overline{u}$ or vice-versa. If found, halt and declare that the given \irc proof cannot be simulated by the algorithm. If no such resolution paths exist, continue to {\bf Step 7}.\\ 
\textbf{Step 7 (Drop universal literals):} Drop all universal literals from all the annotated clauses introduced in Step 2. This completes simulating the axiom download steps of \irc.\\
\textbf{Step 8 (Simulate resolution and instantiation steps):} In the second pass of this algorithm, simulate the resolution and instantiation steps in order as they occur in $\pi$. That is, for every resolution step, add the resolvent clause. Since all the important definition clauses are present, every instantiation steps can be simulated by Lemma~\ref{lemma2}. The completes the algorithm. \\
Let us quickly understand the algorithm with an example. 
\vspace{-0.2cm}
\begin{example}\label{example1}
Consider the following QBF and an \irc proof of the same in Figure~\ref{fig:subfig1}. Apply the modified algorithm on the same.
\vspace{-0.8cm}
\begin{center}
\begin{equation*}
\begin{aligned}
\Psi_0 = {} & \forall u_1 \exists e_2 \forall u_3 \exists e_4,e_5.~(\overline{u_1} \lor \overline{e_2} \lor \overline{u_3} \lor e_5) \land (\overline{u_1} \lor \overline{u_3} \lor \overline{e_4}) \\
&\land ( e_2 \lor \overline{u_3} \lor e_4) \land (u_1 \lor \overline{e_2}) \land (\overline{u_1} \lor \overline{e_2} \lor \overline{e_5})
\end{aligned}
\end{equation*}
\end{center}
\textit{Step 1}: Add definitions for all the annotated literals in $C_1 ,..,C_5$ (Fig.~\ref{fig:subfig1}). There will be a total of 12 definition clauses added. QBF is now $Q^{'}.\Psi_1$\\
\textit{Step 2}: \hspace{9.8cm} Labels
\vspace{-1cm}
\begin{center}
\begin{equation*}
\begin{aligned}
\Psi_2 ={} & \Psi_1 \land (\overline{u_1} \lor \overline{e_2}^{u_1} \lor \overline{u_3} \lor {e_5}^{u_1 u_3}) \land (\overline{u_1} \lor \overline{u_3} \lor \overline{e_4}^{u_1 u_3}) \hspace{1.5cm} (C_1^{'},C_2^{'})\\
& \land ( e_2 \lor \overline{u_3} \lor {e_4}^{u_3} ) \land (u_1 \lor \overline{e_2}^{\overline{u_1}}) \land (\overline{u_1} \lor \overline{e_2}^{u_1} \lor \overline{e_5}^{u_1}) \hspace{1cm} (C_3^{'},C_4^{'},C_5^{'})
\end{aligned}
\end{equation*}
\end{center}
\vspace{-0.2cm}
\textit{Step 3}: \hspace{3cm}$\Psi_3 = \Psi_2 - \Psi_0$\\
\textit{Step 4}:(from \irc proof in Figure \ref{fig:subfig1})\\
1: $C_6 = inst(u_1,C_3)$ \hspace{0.1cm} : \hspace{0.1cm}
Required clauses = $(\overline{e_4}^{u_3} \lor {e_4}^{u_1 u_3}) \land (\overline{e_2} \lor {e_2}^{u_1})$\\
\hspace*{4cm} Imp. Def clauses = $(e_4 \lor \overline{e_4}^{u_3}) \land (\overline{e_4} \lor {e_4}^{u_1 u_3}) \land (\overline{e_2} \lor {e_2}^{u_1})$\\
2: $C_{10} = inst(u_3,C_8)$ \hspace{0.1cm} : \hspace{0.1cm}
Required clauses = $({e_5}^{u_1} \lor \overline{e_5}^{u_1 u_3})$\\
\hspace*{4cm} Imp. Def clauses = $(\overline{e_5} \lor {e_5}^{u_1}) \land (e_5 \lor \overline{e_5}^{u_1 u_3})$\\
\textit{Step 5}: Drop non-important definitions, now formula will be:
\vspace{-0.2cm}
\begin{center}
$\Psi_4 = \Psi_3 - \Psi_1 + \{ (e_4 \lor \overline{e_4}^{u_3}) \land (\overline{e_4} \lor {e_4}^{u_1 u_3}) \land (\overline{e_2} \lor {e_2}^{u_1}) \land (\overline{e_5} \lor {e_5}^{u_1}) \land (e_5 \lor \overline{e_5}^{u_1 u_3}) \}$
\end{center}
\vspace{-0.2cm}
\textit{Step 6}:\\
Rightmost $\forall$ variable = $u_3$  : 
But no opposite literals in any clause pairs.\\
Next rightmost $\forall$ variable = $u_1$  :
No paths found in $\Psi_4$.\\
\textit{Step 7} \& \textit{8}: Drop all $\forall$ variables from $\Psi_4$. Now, resolvent and instantiated clauses of \irc can be directly added in order since they are AT w.r.t the QBF at that point.
This completes the simulation.
\end{example}
\begin{center}
\begin{figure}
\begin{minipage}{5cm}
\begin{subfigure}[h]{5cm}

\begin{tikzpicture}[scale=0.485, transform shape]
\node[ellipse,draw,label=left:{$C_{11}$}] (a) at (5.7, 0) {{\large$\perp$}};
\node[ellipse,draw,label=right:{$C_{10}$}] (b) at (9, 0.6) {{\Large$\overline{e_5}^{u_1 u_3}$}};
\node[ellipse,draw,label={[label distance=-0.1cm]210:$C_9$}] (c) at (2.5, 1.2) {{\Large${e_5}^{u_1 u_3}$}};
\node[ellipse,draw,label={[label distance=-0.2cm]45:$C_8$}] (d) at (9, 2.2) {{\Large$\overline{e_5}^{u_1}$}};
\node[ellipse,draw,label=below:{$C_7$}] (e) at (4.8, 2.6) {{\Large${e_2}^{u_1}$}};
\node[ellipse,draw,label={[label distance=-0.1cm]left:$C_6$}] (f) at (6.5, 3.6) {{\large${e_2}^{u_1} \lor {e_4}^{u_1 u_3} $}};

\node[ellipse,draw,label={[label distance=-0.28cm]210:$C_2$}] (g) at (3.3,5.2) {{\large$\overline{e_4}^{u_1 u_3}$}};
\node[ellipse,draw,label={[label distance=-0.1cm]left:$C_3$}] (h) at (6.5,5.2) {{\large$e_2 \lor {e_4}^{u_3}$}};
\node[ellipse,draw,label={[label distance=-0.1cm]below:$C_4$}] (i) at (8.85,5.2) {{\large$\overline{e_2}^{\overline{u_1}}$}};
\node[ellipse,draw,label={[label distance=-0.1cm]330:$C_5$}] (j) at (11.5,5.2) {{\large$\overline{e_2}^{u_1} \lor \overline{e_5}^{u_1}$}};
\node[ellipse,draw,label={[label distance=-0.1cm]210:$C_1$}] (k) at (0.1,5.2) {{\large$\overline{e_2}^{u_1} \lor {e_5}^{u_1 u_3}$}};

\node[rectangle,draw,label=above:{${C_2}^{''}$},minimum width = 2.5cm, minimum height = 0.75cm] (l) at (3.5,6.5) {{\large$\overline{u_1} \lor \overline{u_3} \lor \overline{e_4}$}};
\node[rectangle,draw,label=above:{${C_3}^{''}$},minimum width = 2.5cm, minimum height = 0.75cm] (m) at (6.5,6.5) {{\large$e_2 \lor \overline{u_3} \lor e_4$}};
\node[rectangle,draw,label=above:{${C_4}^{''}$},minimum width = 1.5cm, minimum height = 0.75cm] (n) at (9,6.5) {{\large$u_1 \lor \overline{e_2}$}};
\node[rectangle,draw,label=above:{${C_5}^{''}$},minimum width = 3cm, minimum height = 0.75cm] (o) at (11.6,6.5) {{\large$\overline{u_1} \lor \overline{e_2} \lor \overline{e_5}$}};
\node[rectangle,draw,label=above:{${C_1}^{''}$},minimum width = 3cm, minimum height = 0.75cm] (p) at (0,6.5) {{\large$ \overline{u_1} \lor \overline{e_2} \lor \overline{u_3} \lor e_5 $}};

\draw[black, ->] (l) -- (g);
\draw[black, ->] (m) -- (h);
\draw[black, ->] (n) -- (i);
\draw[black, ->] (o) -- (j);
\draw[black, ->] (p) -- (k);
\draw[black,dashed, ->] (h) -- (f)node[pos=.5,above, left] {$u_1$};
\draw[black, ->] (f) -- (e);
\draw[black, ->] (g.south) to [bend right] (e.west);
\draw[black, ->] (e) -- (c);
\draw[black, ->] (k.south) to [bend right] (c.west);
\draw[black, ->] (e) -- (d);
\draw[black, ->] (j.south) to [bend left] (d.east);
\draw[black,dashed, ->] (d) -- (b)node[pos=.5,above, left] {$u_3$};
\draw[black, ->] (b.west) -- (a);
\draw[black, ->] (c) -- (a);
\end{tikzpicture}
\caption{\irc proof of $\Psi_0$ (Example~\ref{example1})}
\label{fig:subfig1}
\end{subfigure}
\end{minipage}
\hspace{2.2cm}
\begin{minipage}{5cm}
\begin{subfigure}[h]{5cm}

\begin{tikzpicture}[scale=0.5, transform shape]
\node[ellipse,draw,label=above:{$C_{12}$}] (a) at (5, 0) {{\large$\perp$}};
\node[ellipse,draw,label={[label distance=-0.1cm]below:$C_{11}$}] (b) at (8.5, 1.3) {{\Large$e_1$}};
\node[ellipse,draw,label={[label distance=-0.1cm]below:$C_{10}$}] (c) at (1.5, 1.3)  {{\large$\overline{e_1}$}};
\node[ellipse,draw,,label={[label distance=-0.1cm]below:$C_9$}] (d) at (7,2.7) {{\Large$e_1 \lor \overline{c_1}^{u_1}$}};
\node[ellipse,draw,label={[label distance=-0.1cm]below:$C_8$}] (e) at (3, 2.7) {{\Large$\overline{e_1} \lor \overline{c_2}^{\overline{u_1}}$}};
\node[ellipse,draw,label={[label distance=-0.1cm]above:$C_7$}] (f) at (7, 4.7) {{\large$e_1 \lor \overline{c_1}$}};
\node[ellipse,draw,label={[label distance=-0.1cm]above:$C_6$}] (g) at (3,4.7) {{\large$\overline{e_1} \lor \overline{c_2}$}};

\node[ellipse,draw,label={[label distance=-0.1cm]below:$C_1$}] (h) at (0,2.7) {{\large${c_2}^{\overline{u_1}}$}};
\node[ellipse,draw,label={[label distance=-0.1cm]left:$C_2$}] (i) at (1.5,6) {{\large$\overline{e_1} \lor c_1$}};
\node[ellipse,draw,label={[label distance=-0.1cm]below:$C_3$}] (j) at (5,6) {{\large$\overline{c_1} \lor \overline{c_2}$}};
\node[ellipse,draw,label={[label distance=-0.1cm]right:$C_4$}] (k) at (8.5,6) {{\large$e_1 \lor c_2$}};
\node[ellipse,draw,label={[label distance=-0.1cm]below:$C_5$}] (l) at (10,2.7) {{\large${c_1}^{u_1}$}};

\node[rectangle,draw,label=above:{${C_1}^{''}$},minimum width = 2.4cm, minimum height = 0.75cm] (m) at (0,4) {{\large$u_1 \lor c_2$}};
\node[rectangle,draw,label=above:{${C_2}^{''}$},minimum width = 2.4cm, minimum height = 0.75cm] (n) at (1.5,7.3) {{\large$\overline{e_1} \lor c_1$}};
\node[rectangle,draw,label=above:{${C_3}^{''}$},minimum width = 2.4cm, minimum height = 0.75cm] (o) at (5,7.3) {{\large$\overline{c_1} \lor \overline{c_2}$}};
\node[rectangle,draw,label=above:{${C_4}^{''}$},minimum width = 2.4cm, minimum height = 0.75cm] (p) at (8.5,7.3) {{\large$e_1 \lor c_2$}};
\node[rectangle,draw,label=above:{${C_5}^{''}$},minimum width = 2.4cm, minimum height = 0.75cm] (q) at (10,4) {{\large$\overline{u_1} \lor c_1 $}};

\draw[black, ->] (m) -- (h);
\draw[black, ->] (n) -- (i);
\draw[black, ->] (o) -- (j);
\draw[black, ->] (p) -- (k);
\draw[black, ->] (q) -- (l);
\draw[black,dashed, ->] (f) -- (d)node[pos=.5,above, left] {$u_1$};
\draw[black, ->] (d) -- (b);
\draw[black, ->] (l) -- (b);
\draw[black, ->] (k) -- (f);
\draw[black, ->] (j) -- (f);
\draw[black, ->] (j) -- (g);
\draw[black, ->] (i) -- (g);
\draw[black, dashed, ->] (g) -- (e)node[pos=.5,above, left] {$\overline{u_1}$};
\draw[black, ->] (e) -- (c);
\draw[black, ->] (h) -- (c);
\draw[black, ->] (b) -- (a);
\draw[black, ->] (c) -- (a);
\end{tikzpicture}
\caption{\irc proof of $\phi_1$}
\label{fig:subfig2}
\end{subfigure}
\end{minipage}
\caption{Example \irc proofs. (Dashed arrow correspond to the instantiation steps).}
\label{fig:subfigureExample}
\end{figure}
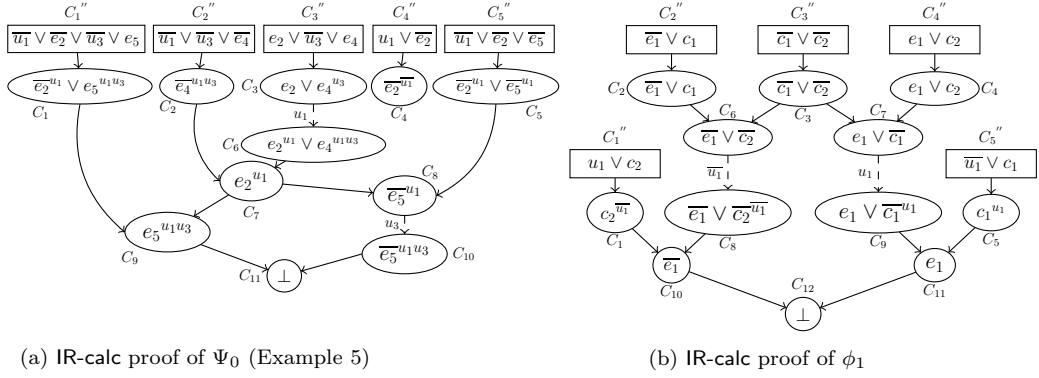
\end{center}
\vspace{-1.5cm}
\section{Counter-Example:}
We show that the proposed two pass algorithm cannot simulate every \irc proof. 
Consider the following false family of QBFs $\phi_n$ from~\cite{exp_paper} and an \irc proof for $\phi_1$ in Figure \ref{fig:subfig2}. Apply the modified algorithm on $\phi_1$.
\vspace{-0.8cm}
\begin{center}
\begin{align}\nonumber
\phi_n	&\equiv \exists e_1\forall u_1 \exists c_1 c_2 \dots \exists e_n \forall u_n \exists c_{2n-1} c_{2n}. \nonumber \\
	& \bigwedge_{i \in [n]} \{ (\overline{e_i} \lor c_{2i-1}) \land (\overline{u_i} \lor c_{2i-1}) \land (e_i \lor c_{2i}) \land (u_i \lor c_{2i}) \} \land (\bigvee_{i \in [2n]} \overline{c_i} )\ \nonumber
\end{align}
\end{center}
\vspace{-1cm}
\begin{center}
\begin{equation*}
Q.\Phi_0 = \phi_1 =\exists e_1\forall u_1 \exists c_1 c_2.~ (u_1 \lor c_2) \land (\overline{e_1} \lor c_1) \land (\overline{c_1} \lor \overline{c_2}) \land (e_1 \lor c_2) \land (\overline{u_1} \lor c_1)
\end{equation*}
\end{center}
\textit{Step 1}:This QBF will need a total of 4 definition clauses. QBF is now $Q^{'}.\Phi_1$.\\
\textit{Step 2}: \hspace{8.8cm} Labels
\vspace{-1cm}
\begin{center}
\begin{equation*}
\begin{aligned}
\Phi_2 ={} & \Phi_1 \land (u_1 \lor {c_2}^{\overline{u_1}}) \land (\overline{e_1} \lor c_1) \land  (\overline{c_1} \lor \overline{c_2}) \hspace{0.7cm} (C_1^{'},C_2^{'},C_3^{'})\\
& \land (e_1 \lor c_2) \land (\overline{u_1} \lor {c_1}^{u_1}) \hspace{3.5cm} (C_4^{'},C_5^{'})
\end{aligned}
\end{equation*}
\end{center}
\textit{Step 3}:\hspace{2.5cm}$\Phi_3 = \Phi_2 - \Phi_0$\\
\textit{Step 4}:(From the \irc proof example in Fig. \ref{fig:subfig2})\\
1: $C_8 = inst(\overline{u_1},C_6)$ \hspace{0.05cm} : \hspace{0.05cm}
Required clauses = $(c_2 \lor \overline{c_2}^{\overline{u_1}})$ = Imp. Def clauses,\\
2: $C_9 = inst(u_1,C_7)$ \hspace{0.05cm} : \hspace{0.05cm}
Required clauses = $(c_1 \lor \overline{c_1}^{u_1})$ = Imp. Def clauses.\\
\textit{Step 5}: \hspace{0.2cm}
$\Phi_4 =\Phi_3 - \Phi_1 + \{ (c_2 \lor \overline{c_2}^{\overline{u_1}}) \land (c_1 \lor \overline{c_1}^{u_1}) \}$ \hspace{0.8cm} $(D_1,D_2)$\\
\textit{Step 6}:\hspace{0.1cm}
Rightmost $\forall$ variable = $u_1$  :
$C_5^{'}$ has $\overline{u_1}$ and $C_1^{'}$ has $u_1$\\
\hspace*{3cm}Resolution path: $C_5^{'},D_2,C_3^{'},D_1,C_1^{'}$\\
We have a path where every clause is important so algorithm fails and halts.
Similarly, the formulas $\phi_n$ cannot be simulated for any $n$.

\section{Discussions and conclusions} 
In this short note, we show that the \qrat simulation algorithm for \ecalculus cannot be lifted to \irc. The only approach to lift this algorithm is similar to the two pass algorithm defined in Section~\ref{sec:modified-algo}. We showed that the modified algorithm cannot simulate the \irc proof of the formula $\phi_n$, which is known to be easy for \irc but hard for \ecalculus. Whether this is always the case is unclear. That is, does the algorithm always fail to simulate the \irc proof of QBFs which are hard for \ecalculus? 
In closing, it is still open `whether \qrat can simulate \irc?'
\bibliographystyle{elsarticle-num} 
\section*{\refname}
\bibliography{myrefs}
\end{document}